\documentclass[12pt]{article}
\usepackage{amsmath}
\usepackage{mathtools}
\usepackage{amssymb}
\usepackage{graphicx}

\title{Generation of light beams with custom orbital angular momentum and tunable transverse intensity symmetries}
\date{June 2019}
\begin{document}
\maketitle
\subsubsection*{Authors:}
Job Mendoza-Hern\'{a}ndez,$^{1,3}$  Mateusz Szatkowski,$^{1,2}$  Manuel F. Ferrer-Garcia,$^{1}$ \\ Julio C. Guti\'{e}rrez-Vega,$^{1}$ and Dorilian Lopez-Mago$^{1,4}$.
\subsubsection*{Addresses:}
$^{1}$Tecnologico de Monterrey, Escuela de Ingenier\'{i}a y Ciencias, Ave. Eugenio Garza Sada 2501, Monterrey, N.L., M\'{e}xico, 64849.\\
$^{2}$Wroc\l{}aw University of Science and Technology, Department of Optics and Photonics, Wybrze\.{z}e Wyspia\'{n}skiego 27, 50-370 Wroc\l{}aw, Poland.\\
$^{3}$ job.mendoza@alumno.buap.mx \\
$^{4}$ dlopezmago@tec.mx

\section{abstract}
We introduce a novel and simple modulation technique to tailor optical beams with a customized amount of orbital angular momentum (OAM). The technique is based on the modulation of the angular spectrum of a seed beam, which allows us to specify in an independent manner the value of OAM and the shape of the resulting beam transverse intensity. We experimentally demonstrate our method by arbitrarily shaping the radial and angular intensity distributions of Bessel and Laguerre-Gauss beams, while their OAM value remains constant. Our experimental results agree with the numerical and theoretical predictions.

\section{Introduction}
Beam shaping studies the techniques to modify either the amplitude or phase of a light beam~\cite{DickeyBook}. Particular attention has been devoted to manipulate the amount of orbital angular momentum (OAM). The OAM of light beams was presented by Allen et al.~\cite{PhysRevA.45.8185} in his seminal paper about Laguerre-Gauss beams. The OAM contained in a Laguerre-Gauss is proportional to the topological charge $\ell$ and is generated by an azimuthal phase distribution of the form $\exp(i\ell\varphi)$. This opened a large number of applications in optical trapping, manipulation, and communications~\cite{torres1,Rubinsztein}. 

There are several methods to generate light beams with OAM. We can use spiral phase plates, superposition of Hermite-Gauss modes, Pancharatnam-Berry phase optical elements (e.g. q-plates), and spatial light modulators (SLM)~\cite{Yao:11}. These methods usually generate light beams containing integer values of OAM. Custom values of OAM, e.g. fractional values~\cite{Berry2004}, have been realized in the optical spectrum using plasmonic vortex elements~\cite{Wang}, whereas in the extreme ultraviolet spectrum through high harmonic generation~\cite{Alex}. An application of fractional OAM is shown by Alexeyev et al.~\cite{Alexeyev:17}, where they study the propagation of a fractional vortex beam  through an optical fiber.

In this work, we introduce a versatile method to tailor the transverse shape of a light beam, while maintaining its OAM fixed at a given value, not necessarily integer. Moreover, we can specify a particular value of OAM, i.e. it is not restricted to integer numbers. Our experimental method finds its theoretical basis on the work by Martinez-Castellanos et al.~\cite{Martinez-Castellanos:15}.

\section{Description of the beam shaping approach}
It has been demonstrated that the $z$ component of the OAM per photon and per unit length of a paraxial beam $U(\mathbf{r},z)$ is given by~\cite{Martinez-Castellanos:15}
\begin{equation}
    J_{z}=-i \frac{\iint |\Tilde{U}_{0}|^{2} \mathcal{A}^{\ast}\partial_{\phi}\mathcal{A}\,\mathrm{d}\textbf{k}}{\iint |\tilde{U}_{0}|^{2}|\mathcal{A}|^{2}\,\mathrm{d}\textbf{k}},\label{Eq:OAMoperator}
\end{equation}
where $\mathbf{r}=(r\cos\varphi,r\sin\varphi)$, $\mathbf{k}=(\rho\cos\phi,\rho\sin\phi)$, $\tilde{U}_{0}$ is the Fourier transform of a seed beam $U_{0}(r,z)$ with null OAM, $\mathcal{A}(\rho,\phi)$ is an algebraic function in Fourier space, $\partial_{\phi}$ is the partial derivative with respect to $\phi$, and $\mathrm{d}\mathbf{k}=\mathrm{d}k_{x}\mathrm{d}k_{y}$. The function $\mathcal{A}$ is the Fourier representation of a general creation operator $\hat{A}(\partial_{x},\partial_{y})$ that acts on the seed beam such that $U=\hat{A} U_{0}$.

Equation~(\ref{Eq:OAMoperator}) can be further simplified for functions of the separable form $\mathcal{A}(\rho,\phi)=R(\rho)\exp[i \Omega(\phi)]$. This condition leads to 
\begin{equation}
J_{z} = \frac{1}{2\pi} \int_{-\pi}^{\pi} \partial_{\phi} \Omega\,\mathrm{d}\phi.
\end{equation}
We also consider that $\Omega$ has the form
\begin{equation}
    \Omega(\phi) = m \phi + \Theta(\phi),\label{Eq:Omegaphi}
\end{equation}
where $\Theta(\phi)$ is a particular phase modulation and $m\in \mathbb{R}$. Expressing $\Omega(\phi)$ in this form yields
\begin{equation}
    J_{z} = m + \frac{1}{2\pi}\int_{-\pi}^{\pi} \partial_{\phi} \Theta\,\mathrm{d}\phi. \label{Eq:Jz}
\end{equation}
For our experiments we set the following phase modulation:
\begin{equation}
    \Theta(\phi) = a\sin(b\phi)^{c}, \label{Eq:THETA}
\end{equation}
where $a,b$, and $c$ are positive real numbers. With this choice for $\Theta$, it turns out that
\begin{equation}
    \frac{1}{2\pi}\int_{-\pi}^{\pi} \partial_{\phi} \Theta\,\mathrm{d}\phi = \frac{1}{2\pi}a\sin(2\pi b)^{c},
\end{equation}
and hence for integer values of $b$, $\int \partial_{\phi} \Theta\,\mathrm{d}\phi=0$ and the content of OAM is exclusively determined by $m$. Therefore, the phase modulation with $b\in \mathbb{Z}$ provides a degree of freedom to shape the angular symmetry of $U$ without altering its OAM. Furthermore, the seed beam $U_{0}$ and the radial function $R(\rho)$ provide control over the radial symmetry. For example, according to the seed beam, if $R(\rho)=1$ and $\Theta(\phi)=0$, we can produce fractional- or integer-order Laguerre-Gauss and Bessel beams.

\begin{figure}[htbp]
\centering
\includegraphics[width=10 cm]{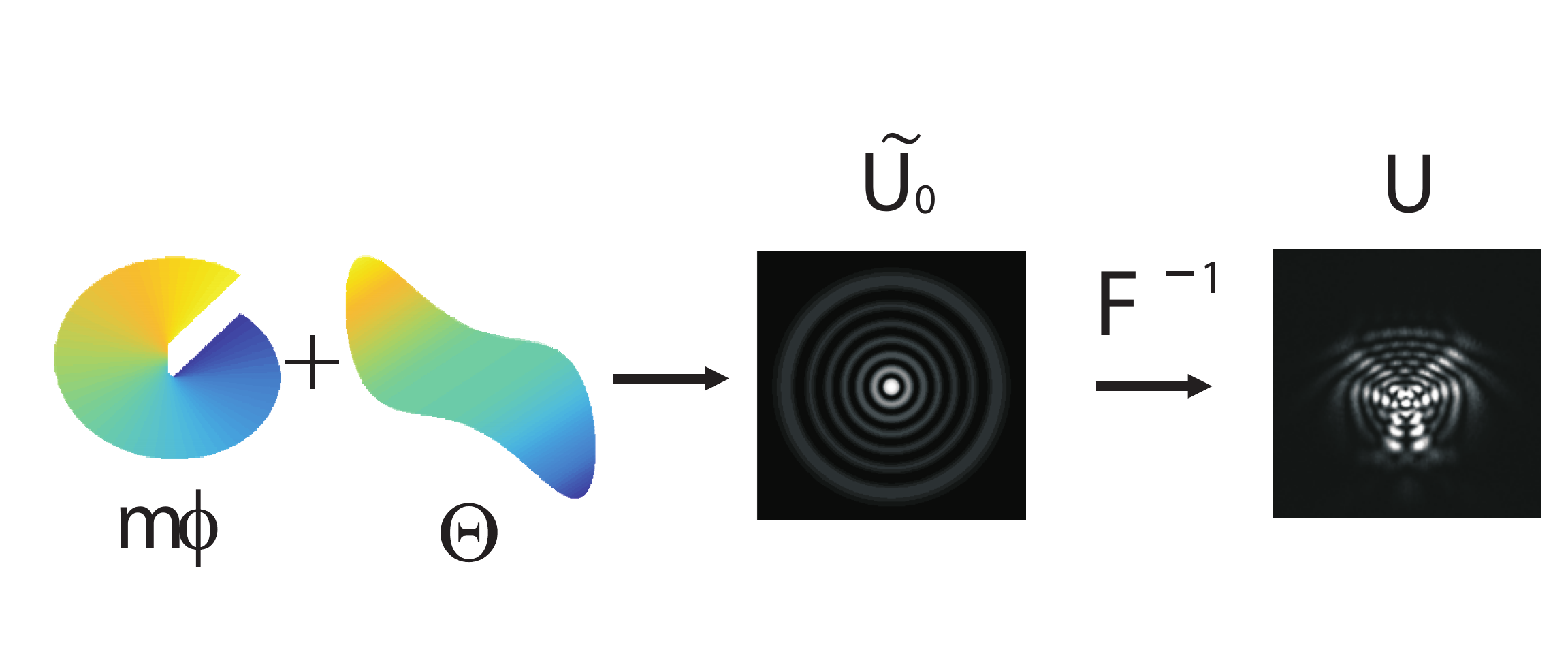}
\caption{(Color online). Intuitive interpretation of the modulation process. The phase modulation $\Omega$ in Eq.~(\ref{Eq:Omegaphi}) is equivalent to the superposition of a spiral phase with a wavefront $\Theta$. The spiral phase determines the value of $J_{z}$ whereas $\Theta$ shapes the intensity of the resulting beam $U$, which results from the diffraction of $\Omega$ with the angular spectrum of the seed beam.}
\label{Fig:1}
\end{figure}

There is an intuitive interpretation of $U=F^{-1}[\mathcal{A}\Tilde{U}_{0}]$ in combination with $\Omega=m\phi + \Theta$ in Eq.~(\ref{Eq:Omegaphi}). The resulting beam $U$ can be seen as the result of $\Omega$ diffracted by the angular spectrum of the seed beam. The value of OAM is given by the spiral phase $m\phi$, whereas the shape of $|U|^{2}$ is determined by the symmetries of $\Theta$. Notice that $m\in \mathbb{R}$ is not necessarily an integer number.  This interpretation is illustrated in Fig.~\ref{Fig:1}.

\section{Experimental arrangement and results}
\begin{figure}[htbp]
\centering
\includegraphics[width=10 cm]{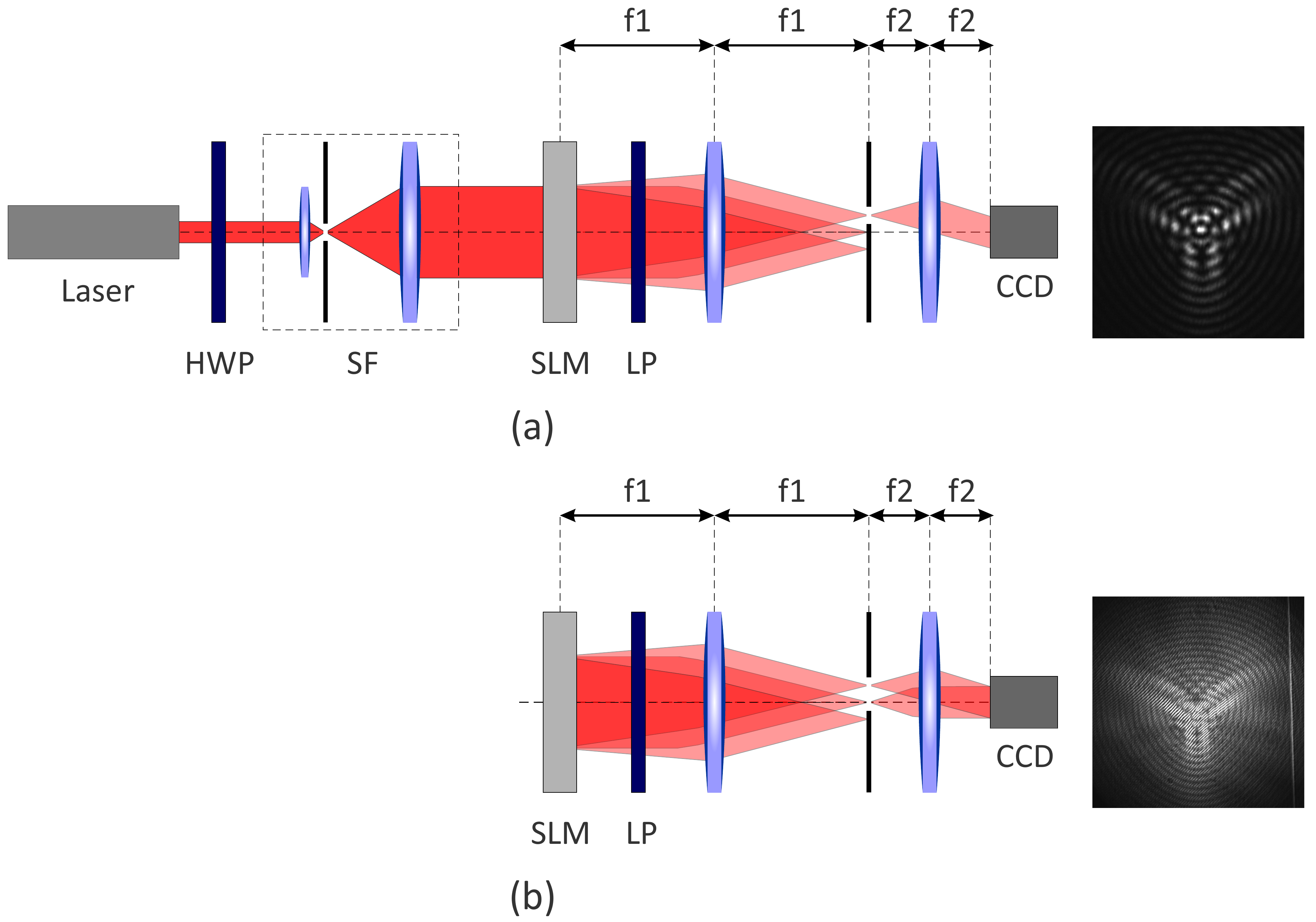}
\caption{Scheme of the experimental setup to generate the structured beam.  (a) detection scheme and (b) phase retrieval scheme. Laser: He-Ne laser with $\lambda=632.8$ nm, HWP: half-wave plate, SF: spatial filter, LP: linear polarizer, SLM: spatial light modulator, lenses with focal lengths $f_{1}=15$ cm and $f_{2}=5$ cm, CCD: camera.}
\label{Fig:2}
\end{figure}

Figure~\ref{Fig:2} shows a schematic of the experimental setup. We apply the technique developed by Arrizon et al.~\cite{Arrizon:05}, which uses an amplitude-only liquid crystal spatial light modulator (SLM) to generate arbitrary complex fields. Given the algebraic function $\mathcal{A}$, the beam amplitude $U$ is calculated with the inverse Fourier transform of $\mathcal{A}\Tilde{U}_{0}$, i.e. $U=F^{-1}[\mathcal{A}\Tilde{U}_{0}]$ (cf.~\cite{Martinez-Castellanos:15}). We use a collimated and linearly polarized He-Ne laser ($632.8$ nm) and a transmissive SLM (HOLOEYE model LC2002). The spatial filter  system (SF) before the SLM cleans the beam transverse intensity and creates a uniform Gaussian beam. The beam polarization is oriented parallel to the SLM director axis using a half-wave plate (HWP). A linear polarizer (LP) is placed after the SLM to eliminate the unmodulated polarization component. The beam $U$ is encoded in a computer generated hologram (CGH) which is displayed in the SLM. A combination of a 4f system with an aperture is implemented to decode $U$ from the first diffraction order, as shown in Fig.~\ref{Fig:2}(a). The intensity pattern $|U|^{2}$ is recorded using a CCD camera. The phase information, i.e. $\arg (U)$, is obtained from the interference pattern between the first and zeroth diffraction orders, as shown in Fig.~\ref{Fig:2}(b). We implemented a phase retrieval algorithm following the work by Takeda et al.~\cite{Takeda:82}.

In the examples that follow, we use three types of seed beams: a Gaussian beam (GB) $\sim \exp[-(2r/w_{0})^{2}]$, a Laguerre-Gauss beam (LGB)
\begin{equation}
    \mathrm{LG}_{n}^{0} = \sqrt{\frac{2}{\pi}} \frac{1}{w_{0}}\exp\left( -\frac{r^{2}}{w_{0}^{2}}\right)\mathrm{L}_{n}^{0}\left( \frac{2r^{2}}{w_{0}^{2}} \right),\label{Eq:LG}
\end{equation}
and a zeroth-order Bessel beam (BB) $\sim J_{0}(k_{t}r)$. In Eq.~(\ref{Eq:LG}), $n$ is the radial index that defines the order of the associated Laguerre polynomial $L_{n}^{0}$ and thus the number of intensity rings. For a fair comparison between the LGB and BB seeds, the transverse wavevector $k_{t}=2\sqrt{2n+1}/w_{0}$ is chosen so that both seed beams have equivalent radii~\cite{Mendoza-Hernandez:15,Mendoza-Hernandez2014}. Furthermore, we choose $R(\rho)=1$ for the LGB and BB seeds, and $R(\rho)=\rho^{2}$ for the GB seed~\cite{Martinez-Castellanos:15}.

\begin{figure}[htbp]
\centering
\includegraphics[width=12 cm]{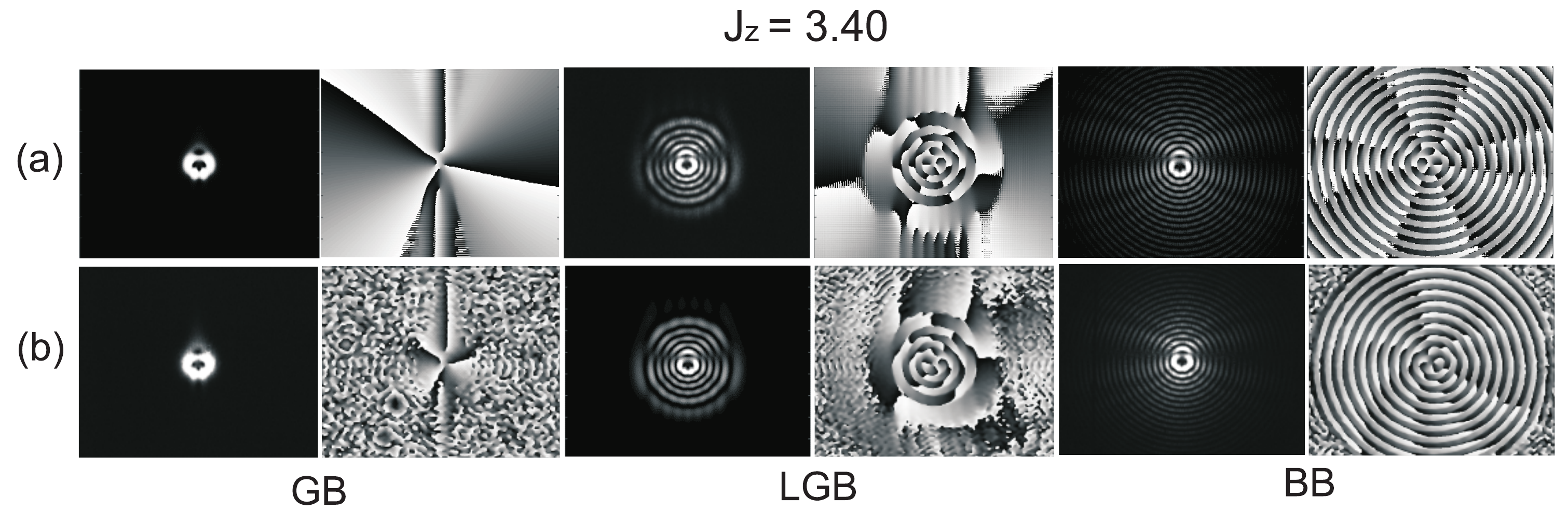}
\caption{Intensity and phase for a structured beam with $J_{z}=3.5$ given by Eq.~(\ref{Eq:Jz}) with $\Theta=0$. GB: Gaussian-type beam due to a Gaussian seed with a radial variation $R(\rho)=\rho^{2}$. LGB: Laguerre-Gauss type beam generated with an LG seed [Eq.~(\ref{Eq:LG})]. BB: Nondiffracting type beam with a zeroth-order Bessel seed. (a) Simulation generated with $U=F^{-1}[\mathcal{A}\Tilde{U}_{0}]$, and (b) experimental results. The numerical value is calculated with Eq.~(\ref{Eq1}) for both the simulation and the experimental measurements. It is equal in both cases.}
\label{Fig:3}
\end{figure}

Figure~\ref{Fig:3} shows the results for the case with $m=3.5$ and $\Theta = 0$ in Eq.~(\ref{Eq:Omegaphi}), which gives $J_{z}=3.5$. Thus, the transverse modulation is defined by the seed beams. First, second and third rows show the resulting beam profiles for the GB, LGB, and BB seeds, respectively. For the LGB case we set $n=5$. The figure shows the simulated beam profile using $U=F^{-1}[\mathcal{A}\Tilde{U}_{0}]$ and the experimental measurements. In both cases we calculate the value of $J_{z}$ using the formal definition~\cite{Padgettbook1}
\begin{equation}
J_{z}=\frac{\iint_{-\infty}^{\infty}\textbf{r}\times \mathrm{Im}(U^{*}\nabla_{\perp}U)\mathrm{d}x\mathrm{d}y}{\iint_{-\infty}^{\infty}|U|^{2}\mathrm{d}x\mathrm{d}y}.\label{Eq1}
\end{equation}
The value of $J_{z}$ for the results shown in Fig.~\ref{Fig:3} turns out to be $3.40$ for both the simulation and experimental measurements, and for all seed beams. The small difference in the value of $J_{z}$, as compared with the theoretical value of $3.5$, is attributed to the discretization of the transverse plane. Nevertheless, there is a good agreement with the theory.

\begin{figure}[htbp]
\centering
\includegraphics[width=12 cm]{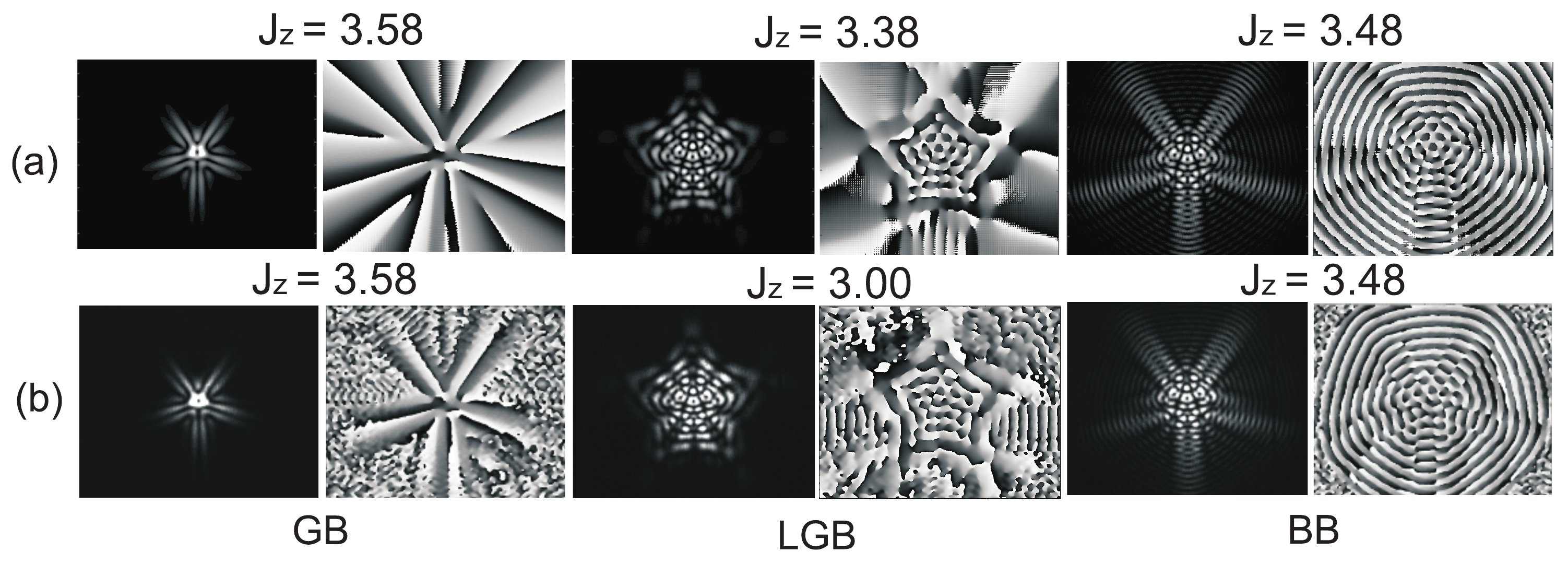}
\caption{Intensity and phase for a structured beam with $J_{z}=3.5$ using Eq.~(\ref{Eq:Jz}) with $\Theta=2\sin(5\phi)$. GB: Gaussian seed. LGB: Laguerre-Gauss seed. BB: Bessel seed. (a) Simulation generated with $U=F^{-1}[\mathcal{A}\Tilde{U}_{0}]$, and (b) experimental results. The numerical values indicate the calculated $J_{z}$ according to Eq.~(\ref{Eq1}).} 
\label{Fig:4}
\end{figure}

Figure~\ref{Fig:4} shows our second example. We set again the value of $J_{z}=3.5$ but we choose $\Theta=2\sin(5\phi)$. The simulations are shown in Fig.~4(a) and the experimental measurements in (b). Notice that the phase modulation $\Theta$ is revealed in the pentagonal symmetry of the intensity pattern. The calculated value of $J_{z}$ for the GB and BB seeds shows excellent agreement between the simulation and the experimental measurements. However, the LGB seed presents a disagreement of about 0.38 between the simulation and the experiment. 

\begin{figure}[htbp]
\centering
\includegraphics[width=12 cm]{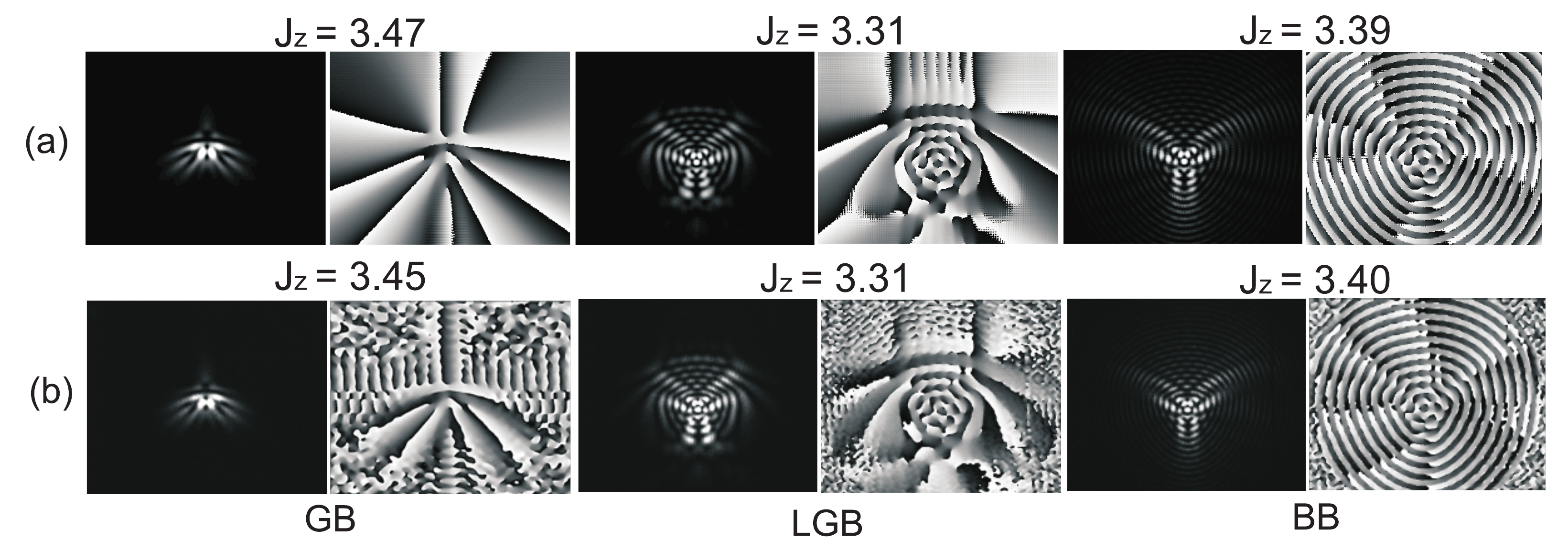}
\caption{Intensity and phase for a structured beam with $J_{z}=3.5$ using Eq.~(\ref{Eq:Jz}) with $\Theta=8\sin(\phi)^{3}$. GB: Gaussian seed. LGB: Laguerre-Gauss seed. BB: Bessel seed. (a) Simulation generated with $U=F^{-1}[\mathcal{A}\Tilde{U}_{0}]$, and (b) experimental results.}
\label{Fig:5}
\end{figure}

Our last example shows a beam profile with a triangular symmetry. It is generated by choosing $\Theta=8 \sin(\phi)^{3}$. Once more, we set the value of $J_{z}$ to $3.5$. The simulation and experimental measurements are shown in Figs.~\ref{Fig:5}(a) and \ref{Fig:5}(b), respectively. Similarly, there is a good agreement between the simulations and experimental values of $J_{z}$.

\section{Conclusions}
We have introduced a method that allows us to control in an independent manner the symmetry of the beam transverse intensity and its amount of OAM. We found a simple expression to modulate the angular spectrum of a seed beam without affecting the OAM. The modulation is given by Eq.~(\ref{Eq:THETA}), which provides the shaping parameters $a,b$ and $c$. From Figs.~\ref{Fig:4} and \ref{Fig:5}, we can see that the shaping parameters $b,c$ determines the polygonal shape of the beam. In general, the shape is determined by the product of $b$ with $c$. The parameter $a$ resembles a focusing factor. For example, in the Fourier domain the operation of a cylindrical lens is given by $\exp (i \sin(\phi)^2/f)$, where $f$ is the focal length. Thus, by comparison, $a\equiv 1/f$ behaves as a focusing parameter. We emphasize that the value of $J_{z}$ is ultimately determined by the seed beam and the algebraic function $\mathcal{A}$, as it is prescribed by Eq.~(\ref{Eq1}). However, Eq.~(\ref{Eq:Jz}) gives a simple expression where it is straightforward to see the relevant parameters that can be modified without affecting the OAM. Our method can find applications in optical tweezers where we can manipulate the particle trajectories according to the beam shape~\cite{Serrano-Trujillo(2018)}.

\section*{Funding}
Consejo Nacional de Ciencia y Tecnolog\'{i}a (CONACYT) (Grants: 257517, 280181, 293471, 295239, APN2016-3140). Polish Ministry of Science and Higher Education ("Diamond Grant") (DIA 2016 0079 45) Nacional Science Centre (Poland) (UMO-2018/28/T/ST2/00125)

\section*{Acknowledgments}
JMH acknowledges partial support from CONACyT, M\'{e}xico. JMH and DLM thank Antonio Morales-Hern\'andez and Alejandra Padilla-Camargo for their help with the SLM.

\bibliography{biblio}

\begin{thebibliography}{10}

\bibitem{DickeyBook}
Fred~M. Dickey.
\newblock {\em Laser Beam Shaping: Theory and Techniques}.
\newblock CRC press, 2014.

\bibitem{PhysRevA.45.8185}
L.~Allen, M.~W. Beijersbergen, R.~J.~C. Spreeuw, and J.~P. Woerdman.
\newblock Orbital angular momentum of light and the transformation of
  laguerre-gaussian laser modes.
\newblock {\em Phys. Rev. A}, 45:8185--8189, Jun 1992.

\bibitem{torres1}
J.~P. Torres and L.~Torner.
\newblock {\em Twisted Photons: Applications of Light with Orbital Angular
  Momentum}.
\newblock Wiley-VCH, 2011.

\bibitem{Rubinsztein}
H.~Rubinsztein-Dunlop, A.~Forbes, M.~V. Berry, M.~R. Dennis, D.~L. Andrews,
  M.~Mansuripur, C.~Denz, C.~Alpmann, P.Banzer, T.~Bauer, E.~Karimi,
  L.~Marrucci, M.~Padgett, M.~Ritsch-Marte, N.~M. Litchinitser, N.~P. Bigelow,
  C.~Rosales-Guzm\'{a}n, A.~Belmonte, J.~P. Torres, T.~W. Neely, M.~Baker,
  R.~Gordon, A.~B. Stilgoe, J.~Romero, A.~G. White, R.~Fickler, A.~E. Willner,
  G.~Xie, B.~McMorran, and A.~M. Weiner.
\newblock Roadmap on structured light.
\newblock {\em J. Optics}, 19:013001, 2017.

\bibitem{Yao:11}
Alison~M. Yao and Miles~J. Padgett.
\newblock Orbital angular momentum: origins, behavior and applications.
\newblock {\em Adv. Opt. Photon.}, 3(2):161--204, Jun 2011.

\bibitem{Berry2004}
M.~V. Berry.
\newblock {Optical vortices evolving from helicoidal integer and fractional
  phase steps}.
\newblock {\em Journal of Optics A: Pure and Applied Optics}, 6:259--268, 2004.

\bibitem{Wang}
Yu~Wang, Peng Zhao, Xue Feng, Yuntao Xu, Fang Liu, Kaiyu Cui, Wei Zhang, and
  Yidong Huang.
\newblock Dynamically sculpturing plasmonic vortices: from integer to
  fractional orbital angular momentum.
\newblock {\em Sci. Rep.}, 6, Nov 2016.

\bibitem{Alex}
Alex Turpin, Laura Rego, Antonio Pic\'on, Julio~San Rom\'an, and Carlos
  Hern\'andez-Garc\'a.
\newblock Extreme ultraviolet fractional orbital angular momentum beams from
  high harmonic generation.
\newblock {\em Sci. Rep.}, 7(43888), 2017.

\bibitem{Alexeyev:17}
C.~N. Alexeyev, A.~O. Kovalyova, A.~F. Rubass, A.~V. Volyar, and M.~A.
  Yavorsky.
\newblock Transmission of fractional topological charges via circular arrays of
  anisotropic fibers.
\newblock {\em Opt. Lett.}, 42(4):783--786, Feb 2017.

\bibitem{Martinez-Castellanos:15}
Israel Martinez-Castellanos and Julio~C. Guti\'{e}rrez-Vega.
\newblock Shaping optical beams with non-integer orbital-angular momentum: a
  generalized differential operator approach.
\newblock {\em Opt. Lett.}, 40(8):1764--1767, 2015.

\bibitem{Arrizon:05}
Victor Arriz\'{o}n, Guadalupe M\'{e}ndez, and David~S\'{a}nchez de~La-Llave.
\newblock Accurate encoding of arbitrary complex fields with amplitude-only
  liquid crystal spatial light modulators.
\newblock {\em Opt. Express}, 13(20):7913--7927, Oct 2005.

\bibitem{Takeda:82}
Mitsuo Takeda, Hideki Ina, and Seiji Kobayashi.
\newblock Fourier-transform method of fringe-pattern analysis for
  computer-based topography and interferometry.
\newblock {\em J. Opt. Soc. Am.}, 72(1):156--160, Jan 1982.

\bibitem{Mendoza-Hernandez:15}
Job Mendoza-Hern\'{a}ndez, Maximino~Luis Arroyo-Carrasco, Marcelo~David
  Iturbe-Castillo, and Sabino Ch\'{a}vez-Cerda.
\newblock Laguerre-gauss beams versus bessel beams showdown: peer comparison.
\newblock {\em Opt. Lett.}, 40(16):3739--3742, Aug 2015.

\bibitem{Mendoza-Hernandez2014}
J.~Mendoza-Hern\'{a}ndez, M.L.~Arroyo Carrasco, M.M.~M\'{e}ndez Otero,
  S.~Ch\'{a}vez-Cerda, and M.D.~Iturbe Castillo.
\newblock New asymmetric propagation invariant beams obtained by amplitude and
  phase modulation in frequency space.
\newblock {\em Journal of Modern Optics}, 61:S46--S56, 2014.

\bibitem{Padgettbook1}
L.~Allen, Stephen~M. Barnett, and Miles~J. Padgett.
\newblock {\em Optical Angular Momentum}.
\newblock Engineering Technology, Physical Sciences, 2003.

\bibitem{Serrano-Trujillo(2018)}
Alejandra Serrano-Trujillo and V\'{i}ctor~Ruiz Cort\'{e}s.
\newblock Orbital angular momentum: origins, behavior and applications.
\newblock {\em Proc. SPIE}, 10744:1074404, 2018.

\end{thebibliography}
\bibliographystyle{unsrt}

\end{document}